\documentclass[
aps,%
12pt,%
final,%
notitlepage,%
oneside,%
onecolumn,%
nobibnotes,%
nofootinbib,%
superscriptaddress,%
noshowpacs,%
centertags]%
{revtex4}

\begin{document}
\selectlanguage{english}

\thispagestyle{empty}

\begin{center}
{\large \bf A new break near 10 TeV in the energy spectrum of protons according to data from space-based instruments:\newline astrophysical interpretation}\bigskip

{\bf \copyright{} 2023 A.~A.~Lagutin$^{\pmb{1)}^*}$ and N.~V.~Volkov$^{\pmb{1)}^{**}}$}\blfootnote{$^{1)}$Altai State University, Radiophysics and Theoretical Physics Department, Barnaul, Russia.}\blfootnote{\;$^*$E-mail: {\tt lagutin@theory.asu.ru}}\blfootnote{$\!^{**}$E-mail: {\tt volkov@theory.asu.ru}}

\begin{quotation}
\noindent{\bf Abstract---}Recent experimental data from space-based instruments of the DAMPE and CALET collaborations have shown that the energy spectrum of protons has a new feature, a break in the $\sim 10$~TeV region. In this energy range, the spectrum index of the observed particles varies from $-2.6$ to $-2.9$.

The purpose of this work is to establish the local sources's position and age that determine this break, the index of the proton generation spectrum in them, as well as the astrophysical interpretation of the results obtained in the DAMPE and CALET experiments.

Within the framework of the model of nonclassical diffusion of cosmic rays developed by the authors, which has break due to the propagation of particles in a sharply inhomogeneous (fractal type) galactic medium, it is shown that break in this energy range is formed by tevatron located at a distance of $\sim 120$~pc from the Earth. These source, whose age is $\sim 5 \cdot 10^5$~years, generate particles with a spectrum index $\sim 2.7$.

The power-law behavior of the proton spectrum before and after the break, soft spectrum of particles generation in the source, first obtained in the DAMPE and CALET experiments, should be considered as an indication of the need to revise the standard paradigm accepted today about the sources of cosmic rays, mechanisms of particle acceleration in them and particles propagation in the Galaxy.
\end{quotation}
\end{center}

\section{Introduction}

The most studied topics in high-energy astrophysics is the origin of Galactic cosmic rays. It is clear that in order to construct a theory of the origin of cosmic rays, it is necessary to solve two fundamental problems, the combined results of which should give a physical explanation of the observed spectra: a model of particle acceleration in sources and a model for transport of these particles from their sources to the Earth in a galactic turbulent (fractal type) medium.

Both of these problems today attract the attention of researchers due to the fact that the observed energy spectra cannot be explained within the framework of the standard scenario of the acceleration and propagation of cosmic rays in the Galaxy.

It should be noted, that in accordance with the standard scenario the high-energy nuclei, detected in the Solar system, ought to be produced by Galactic supernova remnants, at least up to $3\cdot 10^{15}$~eV, by diffusive shock acceleration (DSA) mechanism~\cite{Krymskii:1977,Axford:1977,Bell:1978,Berezhko:1999}. This  mechanism predicts power-law rigidity spectra $J \propto R^{- \gamma}$ with slope $\gamma \sim 2.0-2.2$. The subsequent cosmic rays (CRs) transport in the turbulent Galactic magnetic fields is modeled as a diffusion process in homogeneous medium with the diffusion coefficient $D(R) = D_0 {(R/1~\text{GV})}^{\delta}$~\cite{Ginzburg:1964,Berezinskii:1990}.  Under these assumptions, the spectra of primary CRs generated by the global-scale steady state distribution of sources $S(\rr, R)$ in the Galaxy is described by a single power law $J_G \propto R^{- \eta}$ with index $\eta = \gamma + \delta$, which is clearly at odds with the observed hardening of CR hadrons at GV--TV region (see for example~\cite{ATIC:2009,Ahn:2010,Adriani:2011, Aguilar:2015P,Aguilar:2015,Aguilar:2017CO,Aguilar:2018N}).

These new features of the particle spectrum and the energy dependence of the diffusion coefficient, established in~\cite{Aguilar:2018BC}, lead to a contradiction in the parameters of the particle generation spectrum in sources predicted by the acceleration theory and experimental data.

Indeed, the steady state energy spectrum $J_G$ gives a very simple rule to retrieve the average value of injection index $\gamma$ in CR source, frequently used for interpretation of the cosmic rays phenomena in CR community: $\gamma = \eta - \delta$. Since numerous distant sources determine the behavior of the spectrum in the rigidity range of $R \sim\ 1-10^2$~GV, from the experimental data measured in the Solar system, after solar demodulation and removed the contribution of nearby young sources, we get by fitting that the CR spectrum exponent in the Galaxy is $\eta \gtrsim 3.0$ (see our work~\cite{Lagutin:2021}). With $\delta \sim 0.3$ as it follows from the reported B/C ratio~\cite{Aguilar:2018BC} we find that the slope of the CR injection spectrum at the source may be~$\sim 2.7$ or more.

\begin{figure}[ht!]
\includegraphics[width=.8\textwidth]{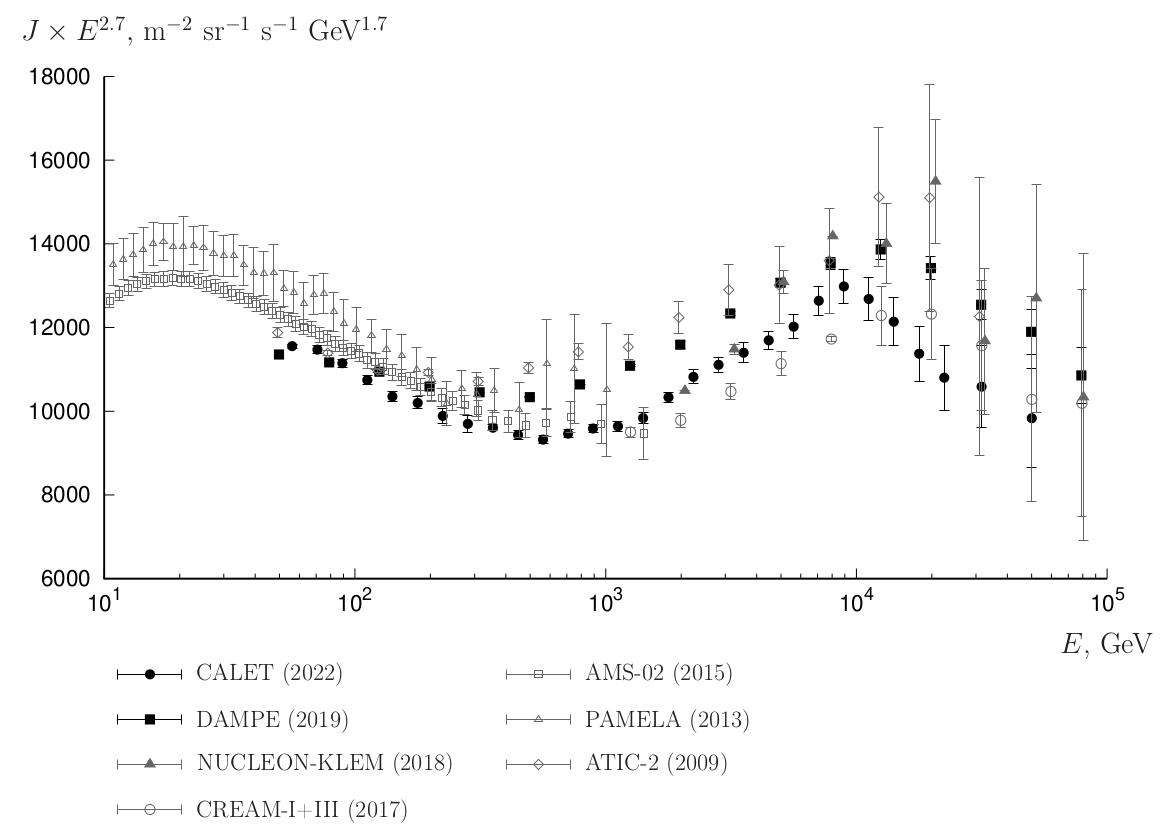}
\caption{The energy spectrum of CR protons observed by CALET~\cite{CALET:2022}, DAMPE~\cite{DAMPE:2019}, NUKLEON-KLEM~\cite{NUCLEON:2018}, CREAM-I+III~\cite{CREAM-III:2017}, AMS-02~\cite{Aguilar:2015P}, PAMELA~\cite{Adriani:2011}, ATIC-2~\cite{ATIC:2009}, multiply by $E^{2.7}$}\label{fig:fig1}
\end{figure}

Recent experimental data from space-based instruments of the DAMPE~\cite{DAMPE:2019} and CALET~\cite{CALET:2022} collaborations confirm the spectral hardening of proton spectra at $\sim$ 300 GeV found by previous experiments and  reveal a softening at $\sim 10$~TeV region (see Fig.~\ref{fig:fig1}). In this energy range, the spectral index of the observed particles varies from $-2.6$ to $-2.9$. A new feature around 10 TeV was previously observed by ATIC~\cite{ATIC:2009}, CREAM~\cite{CREAM-III:2017} and  NUCLEON-KLEM~\cite{NUCLEON:2018}.

In our previous work~\cite{Lagutin:2019} we have shown that the measured by the AMS-02  spectral hardening of the cosmic ray spectra at  $E > 300$~GeV could be caused by the transition from the contribution of multiple distant Galactic sources to the contribution of mainly local ones. However, the lack of experimental data for the range of $\sim 3-100$~TeV at that time did not allow us to evaluate the characteristics of possible tevatrons. The proton spectrum measured by DAMPE up to $\sim 100$~TeV with high statistics allows us to do it today. We also note that the power-law behavior of the proton spectrum before and after the break at the energy of $\sim 10$~TeV, established by DAMPE and CALET, is the main motivating factor in our study, since the CR spectrum in the framework of the nonclassical diffusion model in the limit of low and high energies, below and above the break energy, respectively, has power-law asymptotics (see section 3).

The main goal of this work is to establish the local sources's position and age that determine the spectral hardening and softening at $\sim 10$~TeV, as well as the index of the proton generation spectrum in them.

In the following section we discuss the key elements of nonclassical diffusion model used to describe the cosmic rays propagation in non-homogeneous ISM. In Sections 3 and 4 we present our injection exponent retrieval techniques and findings. Conclusions are given in the last section.

\section{The key elements of nonclassical diffusion model}

The following key assumptions were included in our model to relate the proton spectrum injected by Galactic source with the one measured at the Earth.

\begin{enumerate}
\item Protons are accelerated mainly by the Galactic sources. The spectrum of accelerated particles in the sources is described by the power law  $J\propto E^{-\gamma}$. 

\item The CR sources are divided into two groups as in our paper~\cite{Lagutin:2001CRC}. The first group includes multiple distant ($r\geq 1$~kpc) sources, the second group consists of nearby ($r< 1$~kpc) young ($t< 10^6$~yr) sources. The spatial distribution of sources also suggests the separation of the observed CR fluxes of the nuclei  into two components as follows:
\begin{equation}~\label{JG}
J(\rr,t,E) = J_G(\rr,E)+ \sum\limits_i J_{L_i}(\rr,t,E).
\end{equation}
Here $J_G$ is the global-scale steady state component which was discussed above, $J_{L_i}$ ---  the local component, i.e. the contribution of nearby tevatron.

\item The highly inhomogeneous distribution of matter and magnetic fields in the Galaxy  leads to the nonclassical diffusion of CRs~\cite{Lagutin:2001CRC,Lagutin:2001NP,Lagutin:2003}. Anomalous diffusion is manifested, in particular, by abnormally large free paths of particles (so-called ``L\'{e}vy flights'') with a power-law distribution $p(\rr,E) \propto A(E,\alpha)r^{-\alpha - 1}$, $r \rightarrow \infty$,  $0 < \alpha < 2.$ Besides, a spatially  intermittent magnetic field of the interstellar medium~\cite{Shukurov:2017} results in a higher probability of a long stay of particles in inhomogeneities,  leading to a presence of the so-called ``L\'{e}vy traps''. In the general case, the probability density function $q(t,E)$ of time $t$, during which a particle is trapped in the inhomogeneity, also has a power-law behaviour: $q(t,E) \propto B(E,\beta)t^{-\beta - 1}$, $t \rightarrow \infty$, $\beta < 1$.

\item The equation for the density of particles with energy $E$ at the location $\rr$ and time $t$, generated in highly inhomogeneous medium by Galactic sources with a distribution density $S(\rr,t,E)$ without energy losses and nuclear interactions can be written as~\cite{Lagutin:2001CRC,Lagutin:2003}
\begin{equation}~\label{SuperEq}
\frac{\partial N(\rr,t,E)}{\partial t}=-D(E,\alpha,\beta)\mathrm{D}_{0+}^{1-\beta}(-\Delta)^{\alpha/2} N(\rr,t,E)+ S(\rr,t,E).
\end{equation}
Here $\mathrm{D}_{0+}^{1-\beta}$ denotes the Riemann--Liouville fractional derivative~\cite{Samko:1993} and $(-\Delta)^{\alpha/2}$ is the fractional Laplacian (``Riesz operator'')~\cite{Samko:1993}. The anomalous diffusion coefficient is $D(E,\alpha,\beta) \sim A(E,\alpha)/B(E,\beta) = D_0(\alpha,\beta){(E/1~\text{GeV})}^{\delta}$. In case of $\alpha=2$, $\beta = 1$ we obtain Ginzburg-Syrovatskii's normal diffusion equation.

\item Solution of Eq.~\eqref{SuperEq} for point impulse source $S(\rr,t,E)=S_{0} E^{-\gamma}\delta(\rr) \Theta(T-t)\Theta(t)$ ($\Theta(\tau)$ is the step function) with the power-law injection spectrum and emission time $T$ has the form~\cite{Lagutin:2001CRC,Lagutin:2003}
\begin{equation}\label{eq:solanomdifeq}
N(\rr,t,E)=\frac{S_0 E^{-\gamma}}{D(E,\alpha,\beta)^{3/\alpha}}
\int\limits_{\max[0,t-T]}^{t}d\tau \tau^{-3\beta/\alpha}\Psi_3^{(\alpha,\beta)}\left(|\rr|(D(E,\alpha,\beta)\tau^{\beta})^{-1/\alpha}\right),
\end{equation}
where $\Psi_3^{(\alpha,\,\beta)}(\rho)$ is the density of the fractional stable distribution~\cite{Uchaikin:1999a,Zolotarev:1999}

\begin{figure}[hb!]
\includegraphics[width=.6\textwidth]{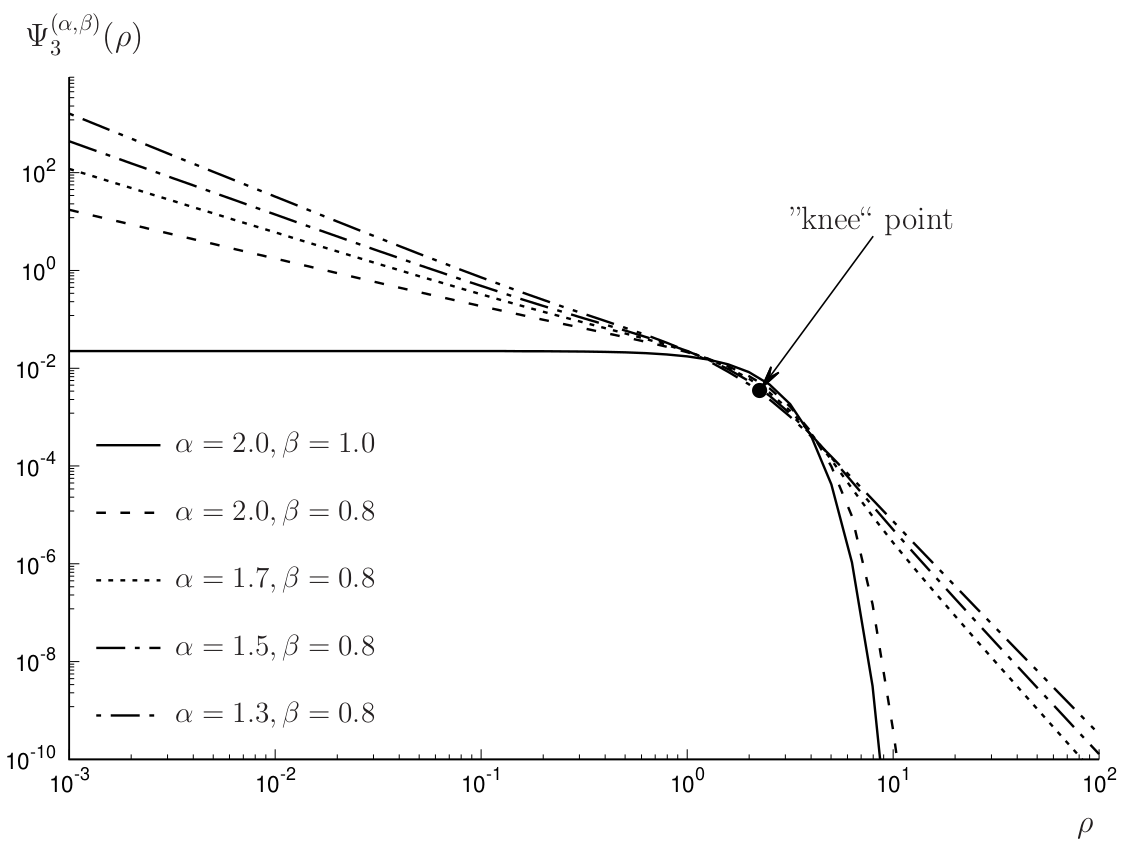}
\caption{Three-dimensional density of fractional stable distribution $\Psi_3^{(\alpha,\beta)}(\rho)$ for different values of~$\alpha$ and $\beta$}\label{fig:fig2}
\end{figure}

\begin{equation}\label{eq:Psi}
  \Psi_3^{(\alpha,\,\beta)}(\rho)=\int\limits_0^\infty{g_3^{(\alpha)}({\rho\tau^\beta})q_1^{(\beta,1)}(\tau)\tau^{3\beta/\alpha}d\tau}.
\vspace*{-3mm}
\end{equation}

Fig.~\ref{fig:fig2} demonstrates the behavior of the function $\Psi_3^{(\alpha,\,\beta)}(\rho)$ for different parameters $\alpha$ and $\beta$. The key feature of this function is the presence of a knee. However, it should be noted that the knee is absent for normal diffusion ($\alpha = 2$, $\beta = 1$)  and subdiffusion ($\alpha = 2$, $\beta < 1$) modes.

\item 
The spectrum of particles from steady state sources is connected with $N(\rr,T,E)$ by means of passage to the limit: $N(\rr,E) = \underset{T \rightarrow \infty}{\lim} N(\rr,T,E)$. In~\cite{Lagutin:2001s} we found that  the global-scale steady state spectrum is
\begin{equation}\label{JG1}
N(\rr,E) \sim E^{-\gamma-\delta/\beta}.
\end{equation}  
It is clear from the physical point of view that the bulk of observed CRs with energy $E \sim 1-10^2$~GeV is formed by numerous distant sources. It means that the CR fluxes in this energy region must be described by Eq.~\eqref{JG1}. 
\end{enumerate}

It should be noted that the model presented above assumes that the diffusion of cosmic rays in a highly inhomogeneous (fractal type) interstellar medium is isotropic. A non-classical model that takes into account the anisotropy of cosmic ray diffusion will be considered in our next work.

\section{ Retrieval of the injection exponent with the use of nonclassical diffusion model  predictions}

Using the representation $N = N_0 E^{-\eta}$ and the property $d\Psi_m^{(\alpha,\,\beta)}(\rho)/d\rho=-2\pi\rho\Psi_{m+2}^{(\alpha,\,\beta)}(\rho)$ of the scaling function $\Psi_3^{(\alpha,\,\beta)}(\rho)$~\cite{Uchaikin:1999a}, one can easily find from~\eqref{eq:solanomdifeq} the spectral exponent $\eta$ for observed particles:
\begin{equation}\label{eq:eta}
\eta = -\frac{d\log N}{d\log R} = \gamma +\frac{\delta}{\alpha}\Xi,
\end{equation}
where
\begin{equation}\label{eq:xi}
\Xi = 3-\frac{2\pi r^2}{D(E,\alpha,\beta)^{2/\alpha}} \frac{\int\limits_{\max[0,t-T]}^{t}d\tau \tau^{-5\beta/\alpha}\Psi_5^{(\alpha,\beta)}\left(|\rr|(D(E,\alpha,\beta)\tau^{\beta})^{-1/\alpha}\right)} {\int\limits_{\max[0,t-T]}^{t}d\tau \tau^{-3\beta/\alpha}\Psi_3^{(\alpha,\beta)}\left(|\rr|(D(E,\alpha,\beta)\tau^{\beta})^{-1/\alpha}\right)}.
\end{equation}

Let $E_k$ be a solution of the equation $\Xi(E_k) = 0$. One can see from~\eqref{eq:eta} and~\eqref{eq:xi} that at $E=E_k$ the spectral exponent for observed particles $\eta$ is equal to spectral exponent for particles generated by the source: $\eta(E_k) = \gamma$. 

Taking into account that  $\rho \equiv |\rr|(D(E,\alpha,\beta)t^{\,\beta})^{-1/\alpha}$, from~\eqref{eq:solanomdifeq} and~\eqref{eq:Psi}  we also find
\begin{equation}\label{Kn:1}
N \sim E^{-\eta} = E^{-\gamma + \delta}, \quad E \ll E_k 
\end{equation}
and
\begin{equation}\label{Kn:2}
N \sim E^{-\eta} = E^{-\gamma - \delta/\beta}, \quad E \gg E_k. 
\end{equation}
Since the exponent $\eta|_{E\ll E_k} = \gamma - \delta$ is less than $\gamma$ at $E\ll E_k$, but the exponent $\eta|_{E\gg E_k} = \gamma + \delta/\beta > \gamma$ at $E\gg R_k$, the value  $E_k$ can be called the knee energy.

Considering that the spectral exponent of protons $\eta$ at the knee energy $E=E_k$, as measured by the DAMPE and CALET experiments, is $\eta(E_k)=2.7$, we find that the injection exponent of particles in local tevatron forming the knee in the $\sim 10$~TeV region is $\gamma =2.7$.

\section{Local sources's position and age}

In the framework of the approach proposed in this paper, the contribution of the local tevatron to proton spectrum can be retrieved from experimental data. The calculation algorithm includes the following steps:
\begin{itemize}

\item subtraction from the observed DAMPE spectrum of the contribution from the system of steady-state sources ($J_G$ component) in the region of $300-10^4$~GeV;

\item subtracting the pevatrons contribution, that affect the behavior of the protons spectrum in the region after the break $10^4-10^5$~GeV;

\item fitting the resulting spectrum with power-law functions before and after the break, establishing the asymptotics of the spectrum of tevatron, determining the spectral index $\eta$;

\item retrieval of the injection index of protons in this tevatron using the equality $\gamma = \eta + \delta$ following from the result~\eqref{Kn:1}.
\end{itemize}

\begin{figure}[ht!]
\includegraphics[width=.8\textwidth]{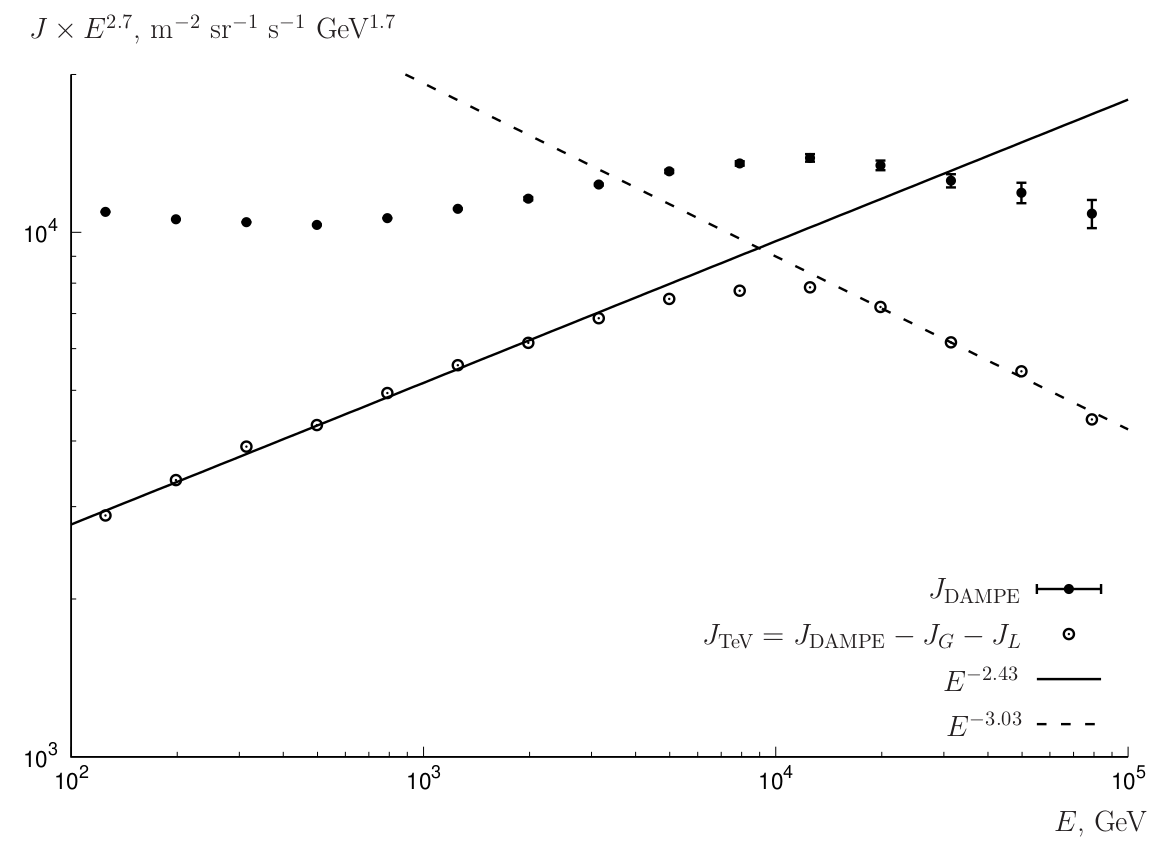}
\caption{Spectra of protons observed by DAMPE~\cite{DAMPE:2019} (filled points). Open points are the local tevatron spectrum $J_{\text{TeV}}$, obtained within the nonclassical diffusion model after subtracting from DAMPE spectrum of the contributions of the both global-scale steady state component $J_G$ and pevatrons one $J_L$. Lines on this figure are the power-law approximations of retrieval DAMPE spectrum of tevatron before and after the break}\label{fig:fig3}
\end{figure}

The spectrum of protons in the Solar system from local tevatron $J_{\text{TeV}}$, obtained within the nonclassical diffusion model, and the slopes of the spectrum before and after the break, shown in the Fig.~\ref{fig:fig3}. A set of model parameters adopted in this paper is given in Table~\ref{tab:adparams}.

\begin{center}
\begin{table}[h!]
\centering
\caption{The  nonclassical diffusion model parameters}\label{tab:adparams}
\begin{tabular}{|l|l|}
\hline
\textbf{Parameter} & \textbf{Value} \\
\hline
$\gamma$ & $2.7$ \\
\hline
$\delta$ & $0.27$ \\
\hline
$D_0(\alpha,\beta)$ & $ 1.5\times 10^{-3}$~pc$^{1.7}$yr$^{-0.8}$\\
\hline
$\alpha$ & $1.7$\\
\hline
$\beta$ & $0.8$ \\
\hline
$T$ & $10^4$ yr\\
\hline
\end{tabular}
\end{table}
\end{center}

Since $\eta|_{E\ll E_k} \sim 2.43$ and $\delta=0.27$, we can retrieve the exponent of the proton generation spectrum at the source using the results of the model:
$$
\gamma = \eta|_{E\ll E_k} + \delta =2.7.
$$

\begin{figure}[ht!]
\includegraphics[width=.6\textwidth]{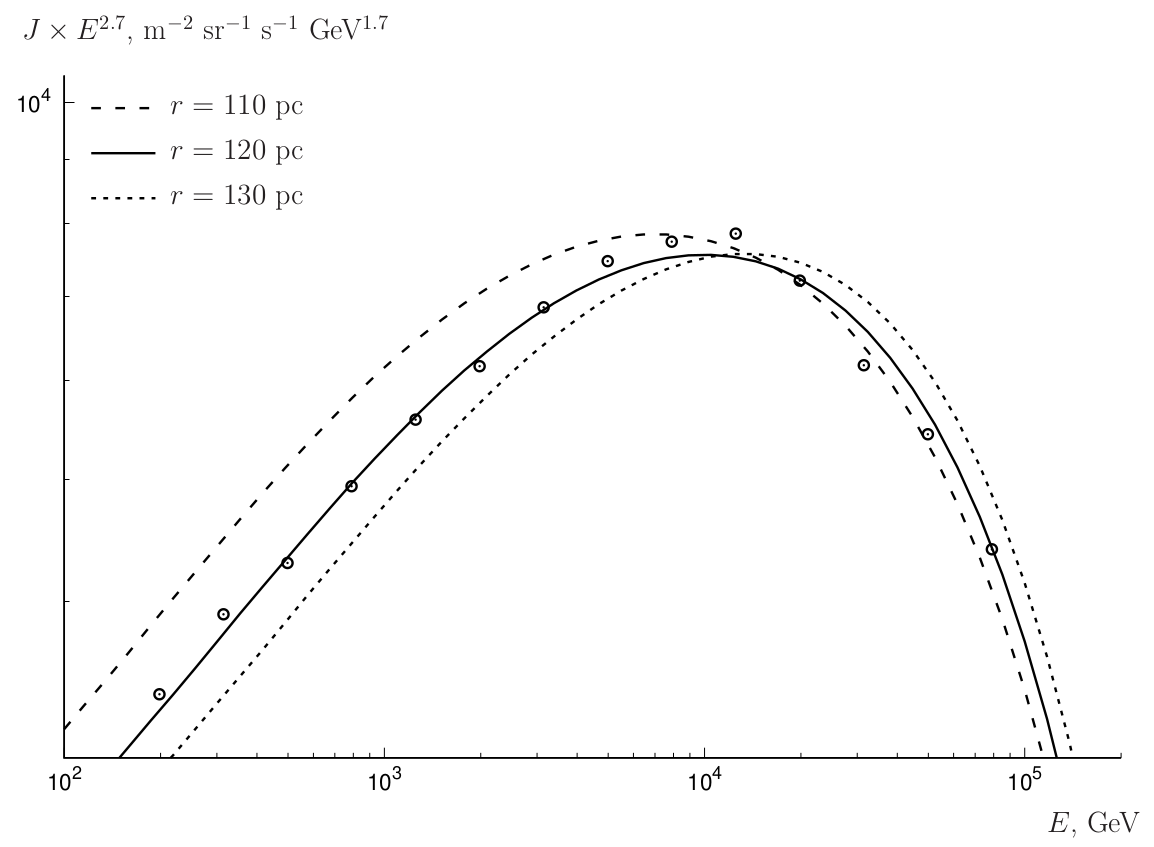}
\caption{Dependence of the proton spectrum on the distance to the tevatron $r$. Points on this figure are the result of retrieval of DAMPE local tevatron spectrum $J_{\text{TeV}}$}\label{fig:fig4}

\includegraphics[width=.6\textwidth]{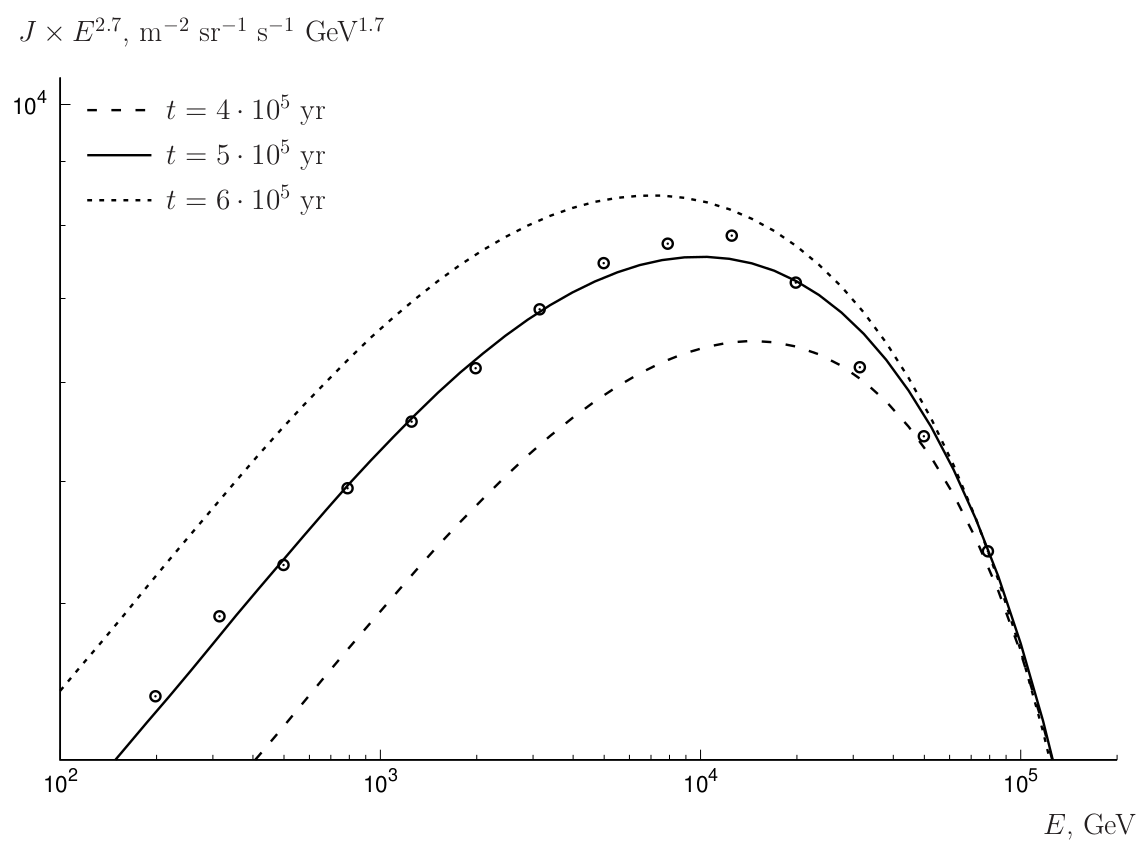}
\caption{Dependence of the proton spectrum on the age of the tevatron $t$. Points in this figure are the same as in Fig.~\ref{fig:fig4}}\label{fig:fig5}
\end{figure}

An analysis of the results of calculations of the proton spectra within the model for point impulse source (see Fig.~\ref{fig:fig4} and~\ref{fig:fig5}) leads to the following estimates:
$$ r \sim 120~{\text {pc}}, \quad t \sim 5 \cdot 10^5~{\text {yr}}.$$
The available estimates of the position and age of the nearest supernovae remnants (see, for example,~\cite{Lozinskaya:1992,Kobayashi:2004}) allow us to conclude that the most likely source that claims to be a tevatron is Loop I.

It should be noted that our estimates of the tevatron's distance and age differ from the results of recent studies~\cite{Kudryashov:2021} and~\cite{Zhao:2023} in which the parameters of the tevatron were obtained using multidimensional optimization and Bayesian inference of the source parameters, respectively.

\section{Conclusion}%

We have presented a novel astrophysical interpretation  of the break around 10 TeV in the energy spectrum of protons. The key results are the following.

\begin{itemize}
\item  The highly inhomogeneous distribution of matter and magnetic fields in the Galaxy  leads to the  non-classical diffusion of CRs. Within the framework of this diffusion model developed by the authors, which has a knee, it is shown that break around 10 TeV is formed by sources located at a distance of 120 pc from the Earth. These sources, whose age is $\sim 5\cdot 10^5$ years, generate particles with a spectral  index $\sim 2.7$. 

\item The power-law behavior of the proton spectrum before and after the break, soft spectrum of particles generation in the source, first obtained in the DAMPE and CALET experiments, should be considered as an indication of the need to revise the standard paradigm accepted today about the sources of cosmic rays, mechanisms of particle acceleration in them and particles propagation in the Galaxy.
\end{itemize}

The mismatch between experimental and theoretical results shown above and spectrum $\sim E^{-2}$ predicted by the standard DSA theory may point to some DSA physics not included in standard treatments~\cite{Bell:2019,Malkov:2019,Blasi:2019}. A radical change in the DSA paradigm for the origin of Galactic CRs is also not ruled out~\cite{Gabici:2019}.

\begin{acknowledgments}
The work is supported by the Russian Science Foundation (grant 23-72-00057).
\end{acknowledgments}

\end{document}